\input{aipcheck}

\documentclass[
    ,final            
  ]
  {aipproc}

\layoutstyle{6x9}

\begin{document}

\title{On the Counter-jet Emission in GRB Afterglows}



\author{Xin Wang and Y. F. Huang}{
  address={Department of Astronomy, Nanjing University, Nanjing 210093, China}
}

\begin{abstract}
We investigate the dynamical evolution of double-sided jets and
present detailed numerical studies on the emission from the receding
jet of gamma-ray bursts. It is found that the receding jet emission
is generally very weak and only manifests as a plateau in the late
time radio afterglow light curves. Additionally, we find that
the effect of synchrotron self-absorption can influence the peak
time of the receding jet emission significantly.
\end{abstract}

\maketitle


\section{Introduction}
According to the popular collapsar progenitor model, a double-sided
jet should be launched by the central engine. However, previously in
calculating the afterglow radiation, people usually only focused on
the emission from the jet component running toward the observer.
Recently the contribution from the jet component running backwards
from us, i.e. the receding jet or counter-jet, was discussed by
several authors \citep{lisong,zhangwq,wang1,wang2}. Especially,
employing the generic dynamical model suggested by
\citet{hqdl2000apj}, we have brought forth detailed numerical
investigations, taking into account many important effects, such
as the effect of equal arrival time surface (EATS), and the effect
of synchrotron self-absorption (SSA), which is decisive in radio
bands \citep{wang1,wang2}.

\section{Numerical results}
For simplicity, we define the following set of parameter values as
the ``standard'' condition: $n=1 / {\rm cm}^{3}$, $E_{\rm 0,iso}
=10^{53} {\rm ergs}$, $\theta_{\rm j} =0.1 $, $\varepsilon = 0 $,
$\xi_{\rm e} = 0.1$, $\xi^2_{\rm B}=0.01$, $p=2.5$, $\theta_{\rm
obs}=0$, $\gamma_0 = 300$, and $z=1$ (corresponding to $d_{\rm
L}= 6634.3$ Mpc).

\begin{figure}
   \includegraphics[width=.33\textwidth]{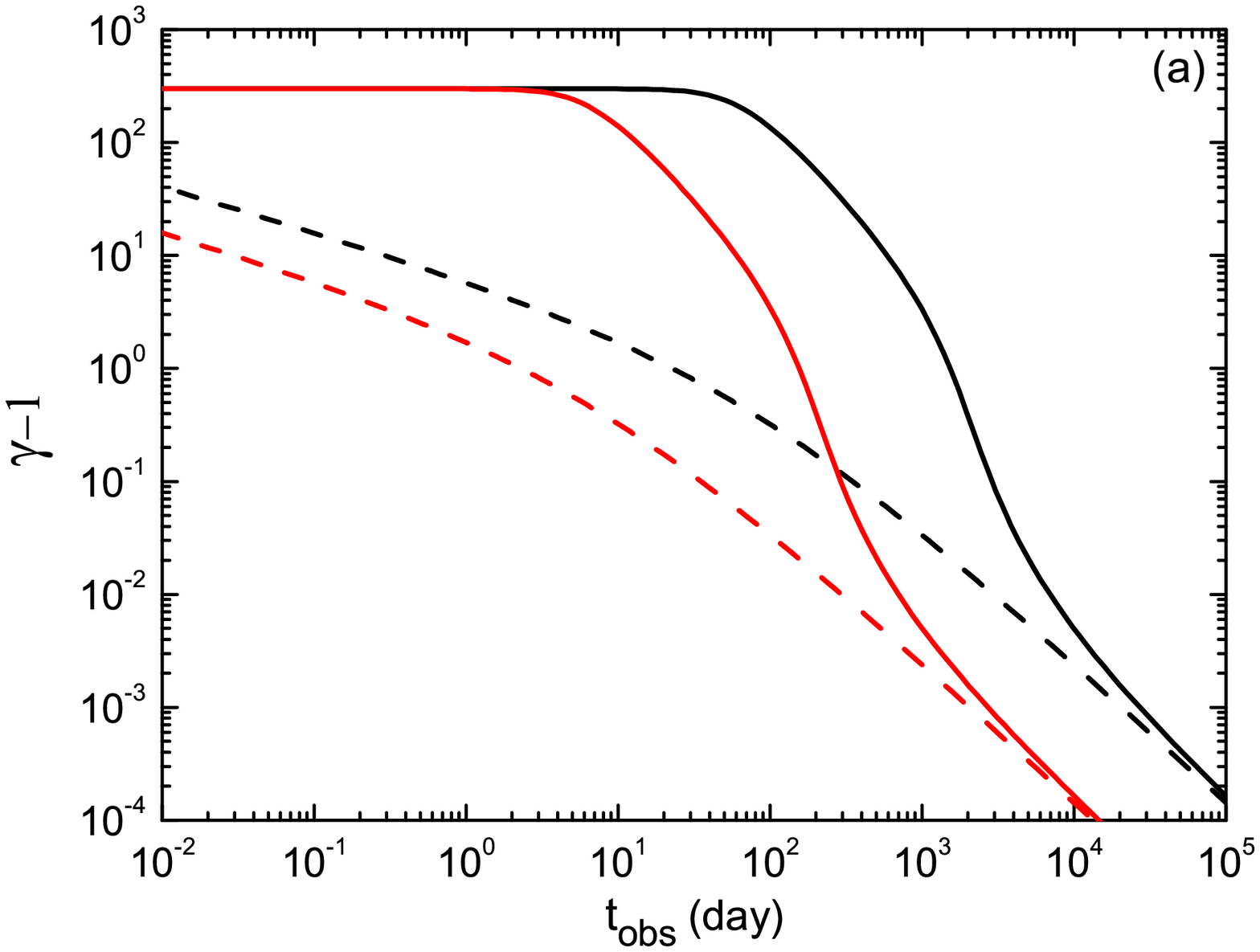} \hspace{-0.2cm}
   \includegraphics[width=.32\textwidth]{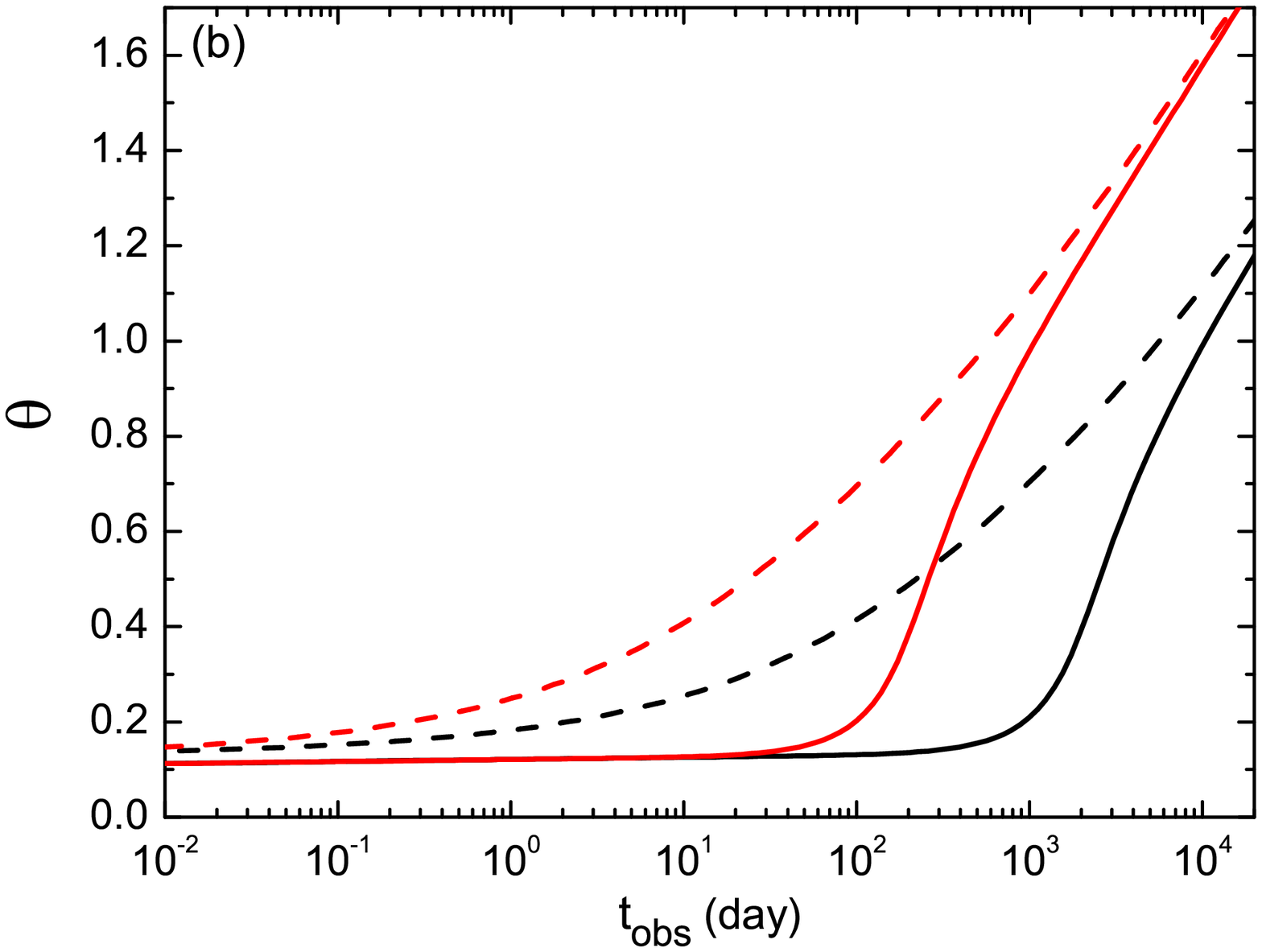} \hspace{-0.2cm}
   \includegraphics[width=.328\textwidth]{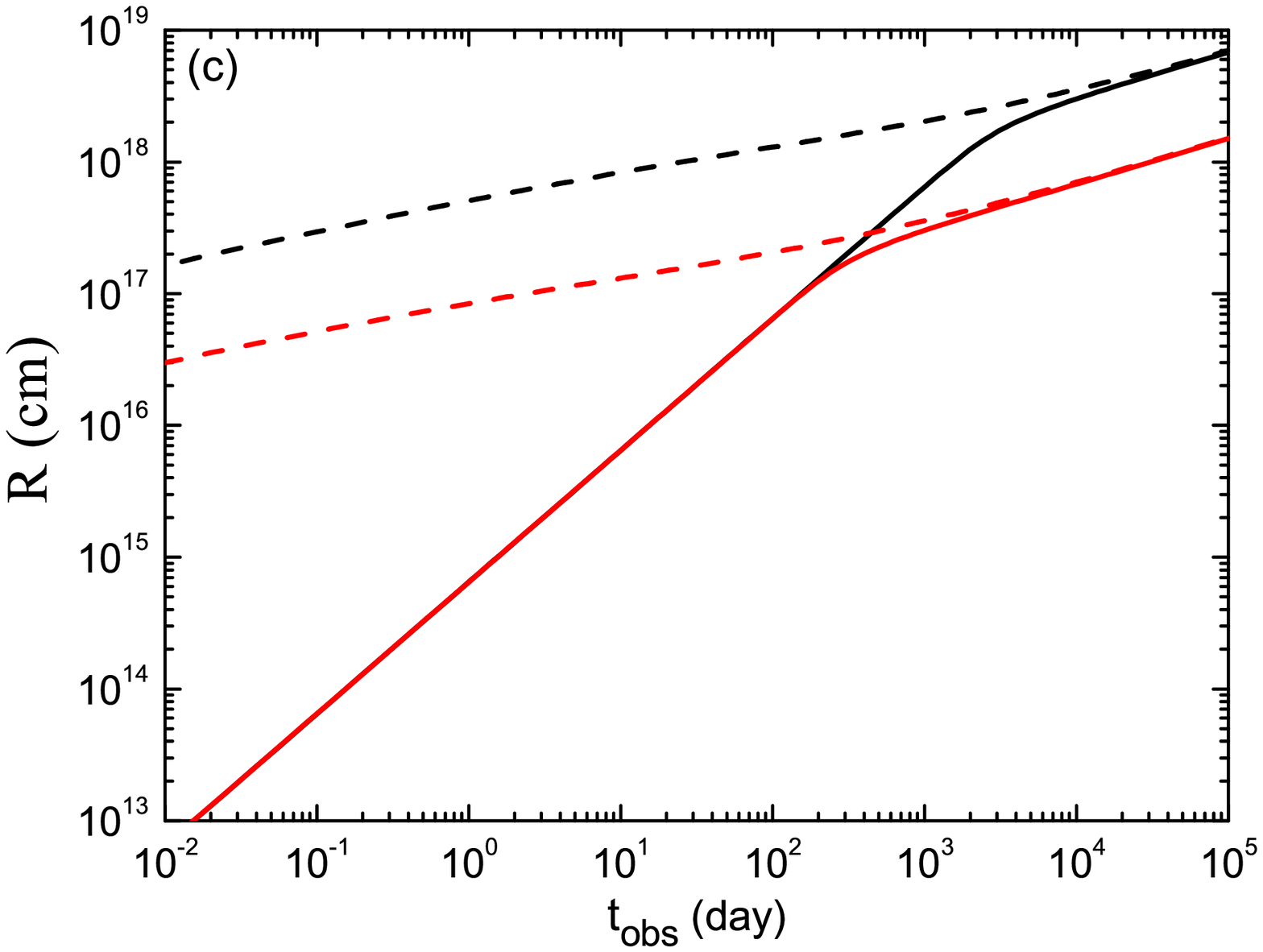}
   \vspace{-0.5cm}
   \caption{The dynamical evolution of the twin jets. In each
   panel, the solid line is plotted for the receding jet while the
   dashed line for the forward jet. The black line set refers to twin
   jets under the ``standard'' condition, and the red line set
   corresponds to the condition by assuming $n=1000 /$cm$^{3}$. The
   observer's time ($t_{\rm obs}$) has been corrected for the
   cosmological effect ($z=1$).}
   \label{fig1}
\end{figure}
In Fig.~1, we illustrate the evolution of three basic physical
quantities, i.e. the Lorentz factors ($\gamma$), the half-opening
angles ($\theta$) and the shock radii ($R$), of the twin jets.
Generally speaking, the physical quantities of the receding jet
remain constant or change mildly in early stage, but evolve rapidly
in a short period afterwards, usually around the time of $10^{2} -
10^{4}$ d. It hints that the relativistic phase for the receding jet
is much longer than that for the forward jet, while the
semi-relativistic stage for the receding jet is much shorter. For
the large circum-burst medium density case ($n=1000 /$cm$^{3}$), the
receding jet is decelerated more rapidly, which means that the
emission from the receding jet will peak earlier than that under the
standard condition.
\begin{figure}
   \includegraphics[width=.7\textwidth]{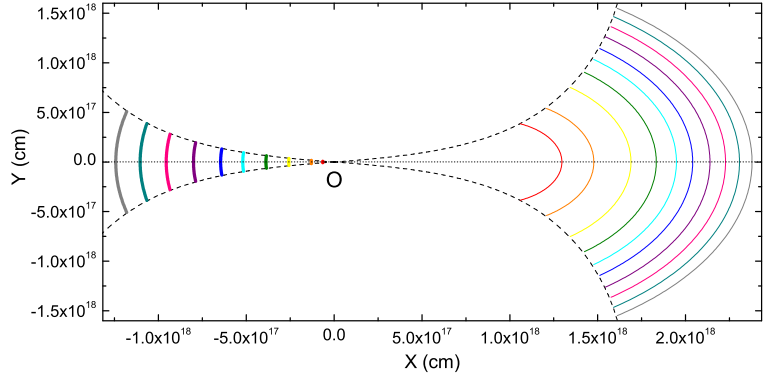}
   \vspace{-0.5cm}
   \caption{Exemplar surfaces for a ``standard'' double-sided jet of ten equal
arrival times, i.e. 100 d (red), 200 d (orange), 400 d (yellow), 600
d (olive), 800 d (cyan), 1000 d (blue), 1250 d (purple), 1500 d
(pink), 1750 d (dark cyan), and 2000 d (grey). ``O'' is the
initiation point of the burst and the dashed lines are the jet
boundaries. The thick solid lines correspond to the EATSs for the
receding jet branch, while thin solid lines are for the forward jet
branch.}
   \label{fig2}
\end{figure}
In Fig.~2, we have shown in total ten exemplars of EATSs. We see
clearly that the EATSs of the receding jet branch have much smaller
typical radius, much flatter curvature and much smaller area, as
compared with those of the EATSs of the forward jet branch, at the
same observer's time.
\begin{figure}
   \includegraphics[width=.5\textwidth]{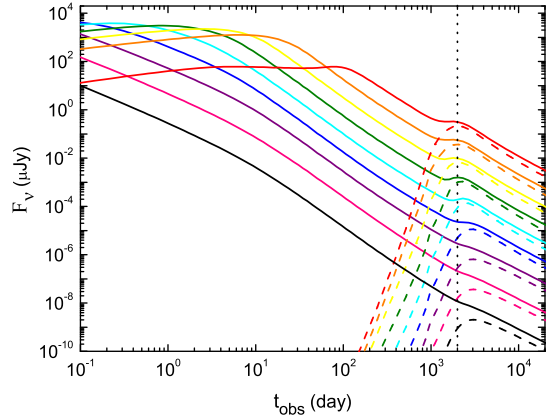}
   \vspace{-0.5cm}
   \caption{Multiband afterglow light curves including the contribution from
the receding jet component. The total light curves at various
observing frequencies, i.e. $10^{9}$ (red), $10^{10}$ (orange),
$10^{11}$ (yellow), $10^{12}$ (olive), $10^{13}$ (cyan), $10^{14}$
(blue), $10^{15}$ (purple), $10^{16}$ (pink), $10^{17}$ (black) Hz,
are represented by solid lines, while the counter-jet emission by
dashed lines. The dotted line marks the peak time of the receding
jet emission at 1 GHz.}
   \label{fig3}
\end{figure}
The total light curves at multiwavelength for a ``standard''
double-sided jet are presented in Fig.~3. Clearly it shows that the
emission from the receding jet really can contribute a significant
portion in the total afterglow light curve at very late stage.
Nevertheless unlike previous work by other authors \citep{lisong}, the
receding jet emission only manifests as a plateau, not an obvious
rebrightening or a marked peak. This discrepancy is mainly ascribed
to the EATS effect, which can be considered only through numerical
calculations \citep{wang1}.

We know that the peak time of the receding jet emission
($t_{\rm{peak}}^{\rm{RJ}}$) is relative to the time when the
receding jet enters the non-relativistic phase
($t_{\rm{NR}}^{\rm{RJ}}$), i.e. $t_{\rm{peak}}^{\rm{RJ}}$ is mainly
determined by dynamics \citep{wang1,wang2}. Therefore it should be
insensitive to the observing frequency (i.e. achromatic). However
Fig.~3 shows that $t_{\rm{peak}}^{\rm{RJ}}$ does not remain constant
over a wide range of frequency, i.e. from radio to optical and
X-ray. This is due to the SSA effect \citep{wang2}, which postpones
the peak time of the receding jet emission and reduces the peak
flux.

\section{Conclusion and discussion}

We have studied the dynamical evolution of double-sided jets and
shown that usually the emission from the receding jet makes a
plateau in the late time afterglow light curves, especially in radio
bands. The flux level of the plateau is usually much less
than 0.3 $\mu$Jy at 1 GHz for the ``standard'' condition. Hence
currently, the counter-jet emission is actually very difficult to
observe. However, the contribution from the receding jet can be
greatly enhanced if the circum-burst environment is very dense
and/or the micro-physics parameters of the receding jet is different
and/or the burst has a low redshift \citep{wang1,wang2}. In these
special cases, if the host galaxy is not very bright as well, then
we may be able to successfully detect the emission from the receding
jet.


\begin{theacknowledgments}
This work was supported by the National Natural Science Foundation
of China (grant 10625313), and by the National Basic Research
Program of China (grant 2009CB824800), and by 2008' National
Undergraduate Innovation Program of China (grant 081028441).
\end{theacknowledgments}



\bibliographystyle{aipproc}   

\bibliography{xwang}

\IfFileExists{\jobname.bbl}{}
 {\typeout{}
  \typeout{******************************************}
  \typeout{** Please run "bibtex \jobname" to optain}
  \typeout{** the bibliography and then re-run LaTeX}
  \typeout{** twice to fix the references!}
  \typeout{******************************************}
  \typeout{}
 }

\end{document}